\documentclass[%
reprint,
superscriptaddress,
 amsmath,amssymb,
 aps,
prx,
floatfix,
]{revtex4-2}

\usepackage{gensymb}
\usepackage{graphicx}
\usepackage{bm}
\usepackage{subfiles}
\usepackage{comment}
\usepackage{hyperref}

\hypersetup{
    colorlinks=true,
    linkcolor=blue,
    citecolor=blue, 
    urlcolor=black
    }

\begin{document}
\setlength{\unitlength}{1cm} 

\title{Measurements of Extended Magnetic Fields in Laser-Solid Interaction}

\author{J. Griff-McMahon}
\email{Corresponding author: jgriffmc@pppl.gov}
\affiliation{Department of Astrophysical Sciences, Princeton University, Princeton, New Jersey 08544, USA}
\affiliation{Princeton Plasma Physics Laboratory, Princeton, New Jersey 08540, USA}

\author{S. Malko}
\affiliation{Princeton Plasma Physics Laboratory, Princeton, New Jersey 08540, USA}

\author{V. Valenzuela-Villaseca}
\affiliation{Department of Astrophysical Sciences, Princeton University, Princeton, New Jersey 08544, USA}

\author{C. Walsh}
\affiliation{Lawrence Livermore National Laboratory, Livermore, California 94550, USA}

\author{G. Fiksel}
\affiliation{Center for Ultrafast Optical Science, University of Michigan, Ann Arbor, Michigan 48109, USA}

\author{M. J. Rosenberg}
\affiliation{Laboratory for Laser Energetics, University of Rochester, Rochester, New York 14623, USA}

\author{D. B. Schaeffer}
\affiliation{Department of Physics and Astronomy, University of California Los Angeles, Los Angeles, California 90095, USA}

\author{W. Fox}
\affiliation{Department of Astrophysical Sciences, Princeton University, Princeton, New Jersey 08544, USA}
\affiliation{Princeton Plasma Physics Laboratory, Princeton, New Jersey 08540, USA}


\begin{abstract}
Magnetic fields generated from a laser-foil interaction are measured with high fidelity using a proton radiography scheme with \textit{in situ} x-ray fiducials. In contrast to prior findings under similar experimental conditions, this technique reveals the self-generated, Biermann-battery fields extend beyond the edge of the expanding plasma plume to a radius of over 3.5 mm by $t$=+1.4 ns, a result not captured in state-of-the-art magneto-hydrodynamics simulations. An analysis of two mono-energetic proton populations confirms that proton deflection is dominated by magnetic fields far from the interaction ($>$2 mm) and electric fields are insignificant. Comparisons to prior work suggest a new physics mechanism for the magnetic field generation and transport in laser-solid interactions.
\end{abstract}

\maketitle

\section{Introduction}
The Biermann-battery mechanism is a fundamental process in plasmas known to self-generate magnetic fields \cite{biermann1950ursprung}. When gradients in electron density and electron temperature are misaligned, resultant thermoelectric currents produce magnetic fields ($\partial B/\partial t \propto \nabla T_e \times \nabla n_e $). At a cosmic scale, the Biermann-battery effect is widely believed to have generated weak, protogalactic seed fields that were amplified by dynamo processes to the $\sim$nT levels observed in galaxies today \cite{kulsrud1997protogalactic}. The generation of strong magnetic fields is also pervasive in laser-solid interactions in which the Biermann-battery mechanism self-generates a toroidal magnetic field around the laser spot \cite{stamper1971spontaneous,haines1986magnetic}. 
Such fields have been observed in simulations of hohlraums which impacts plasma heat transport in inertial confinement fusion (ICF) \cite{farmer2017simulation,walsh2017self}. They also contribute to laser-driven experiments of magnetic reconnection \cite{nilson2006magnetic,willingale2010proton,fox2011fast} and instability growth \cite{manuel2012first,fox2013filamentation}, among others.

Over the last two decades, Biermann-battery fields generated on the surface of laser-driven targets have been characterized extensively with experiments using proton radiography and extended magneto-hydrodynamic (MHD) simulations \cite{li2007observation,petrasso2009lorentz,gao2015precision,campbell2022measuring,willingale2010fast}. These experiments use a beam of MeV protons, commonly from a D$^3$He fusion backlighter, to probe the electromagnetic fields from the interaction. As the protons traverse through the field region, they acquire small deflections from the Lorentz force which are used to infer the path-integrated fields in the plasma \cite{kugland2012invited,li2006measuring,davies2023quantitative}. In prior works, the measured magnetic fields qualitatively agreed with each other in terms of field profile (the fields peak near the laser focal spot) and field dynamics (the furthest reaching fields expand at the plasma sound speed of $\sim500$ km$/$s). 

In this work, magnetic fields are measured on the surface of a laser-driven foil and inferred to have significantly greater magnetic extent, flux, and energy than in previous experiments and simulations under similar conditions \cite{li2007observation,petrasso2009lorentz,campbell2022measuring,gao2015precision}. 
Our measurements show that magnetic fields extend beyond the plasma plume generated on the foil's surface. The large field extent is more than twice as large as the conclusions drawn in similar prior works. 
Likewise, the magnetic flux and energy stored in the magnetic field are elevated by factors of 3$\times$ and 5$\times$, respectively. The novel magnetic field measurement is attributed to our improved proton radiography scheme that uses \emph{in situ} x-ray fiducials to more accurately measure proton deflections, especially at the edge of the detector field of view \cite{johnson2022proton,malko2022design}. 
Additionally, we demonstrate electric fields to be non-significant and confirm that magnetic fields are the primary source of deflection far from the interaction ($>$2 mm).
\begin{figure} [b]
	\includegraphics[width=\linewidth]{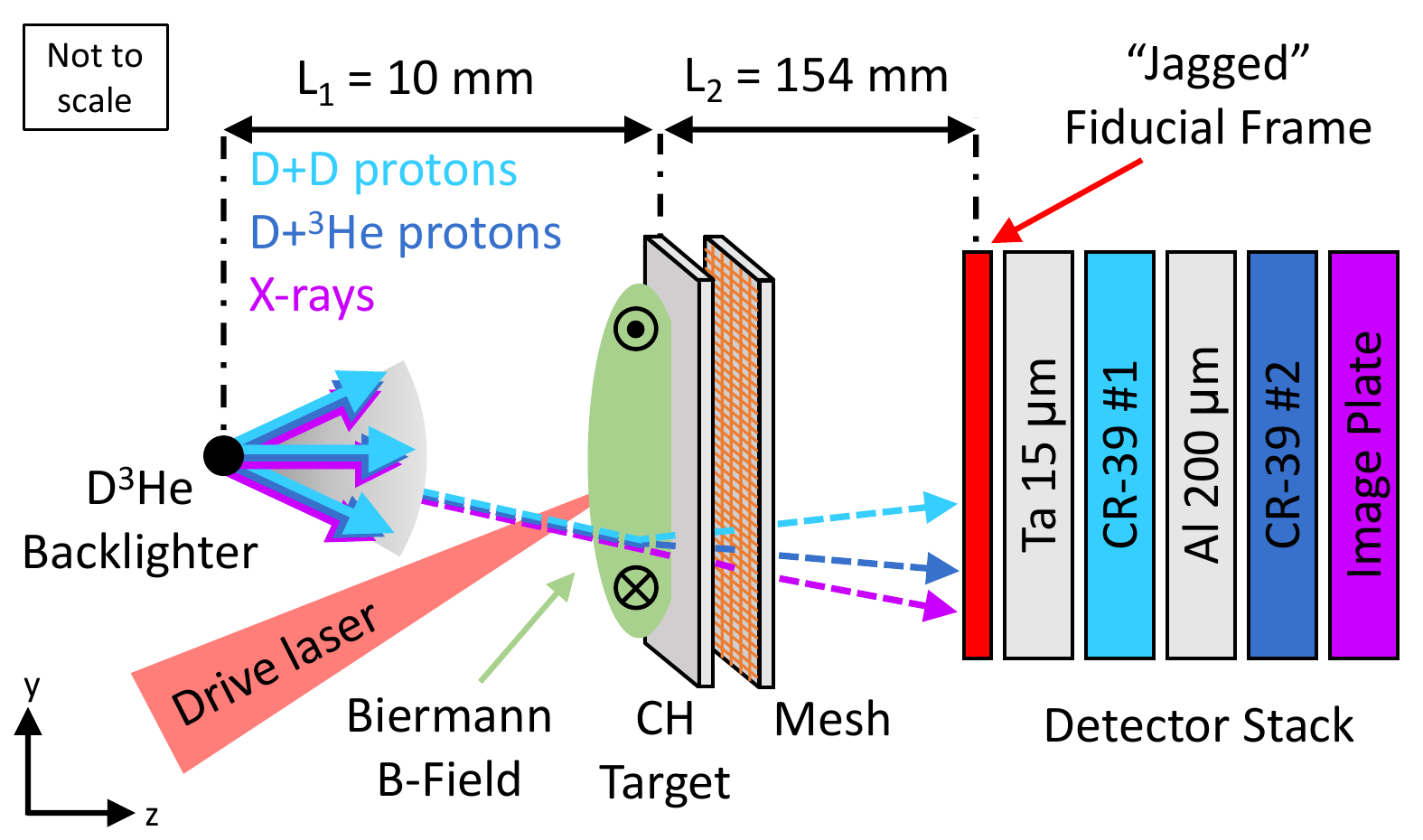}
	\caption{\textbf{Experimental set up.} Biermann-battery magnetic fields are imaged by protons and x-rays. Protons generated from D+D and D+$^3$He fusion reactions have birth-energies of 3 and 14.7 MeV, respectively.}
	\label{fig:ExperimentalSetup}
\end{figure}

\begin{figure*}
    \includegraphics[width=\linewidth]{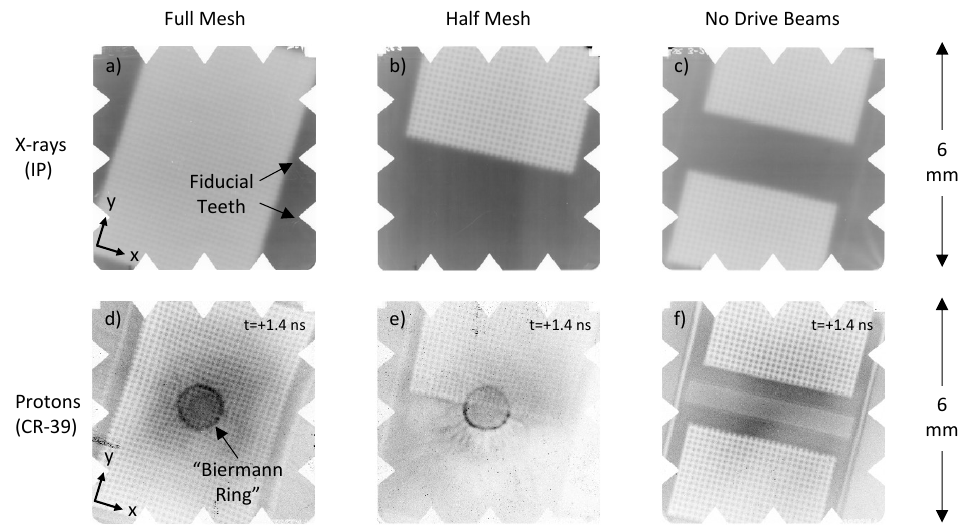}
    \caption{\textbf{Aligned IP and CR-39 images}. The top row (a-c) shows aligned IP images of x-ray fluence and the bottom row (d-f) shows CR-39 images of 14.7 MeV proton fluence (darker regions received higher fluence). A different mesh and beam configuration is shown in each column. The rightmost column has no drive beams and is used as a control shot. Coordinates are in the plasma plane.}
    \label{fig:Combined_IP_CR-39}
\end{figure*}
In proton radiography experiments, a mesh is often used to split the incident proton flux into a grid of “beamlets” to aid in identifying proton deflections. In general, using proton radiography with a mesh yields higher magnetic field accuracy but lower spatial resolution. However, using a mesh and protons alone has a severe limitation; a region of zero beamlet deflection is required to establish the “zero-field” reference pattern of the mesh. 
Critically, if the chosen region contains far-reaching fields but is incorrectly assumed to have zero fields, then the reconstructed reference pattern may be skewed and imply an artificially foreshortened field measurement.

In this experiment, we field an improved proton radiography scheme that also measures x-rays in addition to the usual protons \cite{malko2022design,johnson2022proton}. X-rays emitted by the fusion backlighter are also split into beamlets by a grid, but travel in straight lines to an image plate (IP) at the rear of the detector stack (see Fig. \ref{fig:ExperimentalSetup}). X-ray refraction in the plasma is negligible for characteristic energies of 25 keV recorded on the IP. After aligning the IP to the CR-39 with a fiducial frame, these x-rays give the exact reference pattern of the grid (i.e. fiducial). In a single shot, the absolute path-integrated magnetic field can be inferred from the shift in beamlet position between x-rays and protons as 
\begin{equation}
    \label{eqn:beamlet_def}
    \int d\mathbf{l} \times \mathbf{B} = \frac{m_p v_p \mathbf{d}}{e L_2}
\end{equation}
in the paraxial limit. Here $m_p$ is the proton mass, $v_p$ is the proton velocity, $\mathbf{d}$ is the deflection vector in the detector plane from x-ray to proton beamlet, $e$ is the proton charge, $L_2$ is the distance from the target to the detector, and $d\mathbf{l}$ is the differential path-length along the proton trajectory. This technique measures path-integrated electromagnetic fields at high fidelity and has been benchmarked against vacuum magnetic fields \cite{johnson2022proton}. Crucially, it does not have limiting assumptions that traditional analysis techniques have (i.e. assumption on zero-field reference pattern or assumptions on the proton fluence pattern).

\section{Experimental Setup and Methods}

Experiments were conducted at the OMEGA Laser Facility at the Laboratory for Laser Energetics at the University of Rochester. Magnetic fields were generated by driving a 25 $\mu$m-thick CH foil with two overlapped laser beams. The beams had a 1 ns pulse duration and used an SG5 Distributed Phase Plate \cite{Lin95} with $1/e$ waist of 358 $\mu$m and superguassian exponent of 5.2. The resulting beam spot had a laser intensity of $3\times10^{14}$ W$/$cm$^{2}$. A separate set of 25 beams drove a 400 $\mu$m capsule filled with D$^3$He to produce a backlight source of fusion-product protons and x-rays. The backlighter beams were delayed relative to the drive beams so that 14.7 MeV protons imaged the fields at 1.4 ns after the drive lasers turned on. The protons and x-rays were then split into beamlets from a Ni mesh attached to the rear surface of the target before arriving at the detector stack. The Ni mesh had cell size of 150 $\mu$m and thickness of 60 $\mu$m. A jagged fiducial frame at the front of the detector stack imprinted shadows of the fiducial teeth on the edges of the IP and CR-39 images and helps to align the two images (see Fig. \ref{fig:Combined_IP_CR-39}a).

The IP and CR-39 images are aligned to within 0.4 pixels (130 $\mu$m in image plane), contributing an uncertainty of $0.5$ T$\,$mm of path-integrated magnetic field at 16.4$\times$ magnification \cite{johnson2022proton}. 
Once aligned, pixels in the CR-39 image map directly to pixels in the IP image. Displacement from the x-ray beamlets to the proton beamlets give an absolute measurement of the proton deflection induced by electromagnetic fields. 

Aligned images of the IP and 14.7 MeV proton CR-39 are shown in Figure \ref{fig:Combined_IP_CR-39} with several different mesh and beam configurations that were fielded in the experiment. The ``full mesh'' and ``half mesh'' configurations both have one laser spot, but with mesh covering different portions of the target. The Biermann ``Ring'' feature in the center of these proton images is a common feature in these experiments and approximately corresponds to where the magnetic field strength peaks. The last configuration has no drive beams and is used as a control shot to measure any fields generated from the backlighter or target stalk and quantify the uncertainty in the technique. In addition a ``reconnection'' shot (not shown) was fielded that has two laser spots separated by 1.6 mm with mesh covering the non-reconnecting side of the interaction. The fields generated on the non-reconnecting side should approximate those generated from a single laser spot. 

The beamlet positions were identified using a weighted centroid around points in a regular grid for x-rays and manually for protons. In this experiment, the Biermann ``Ring'' feature is a caustic where two proton beamlets are deflected onto the same final position on the detector. These beamlets are given larger uncertainty in their final position. Most x-ray and proton beamlet locations are identified to within $\pm 1$ pixel (corresponding to 1.2 T$\,$mm), except for a few select proton beamlets that are either in the caustic, have poor signal-to-noise, or have large deviations from their neighboring beamlets, suggesting an inaccurate initial identification of the beamlet.

\section{Results}

\begin{figure} [t]
    \includegraphics[width=\linewidth]{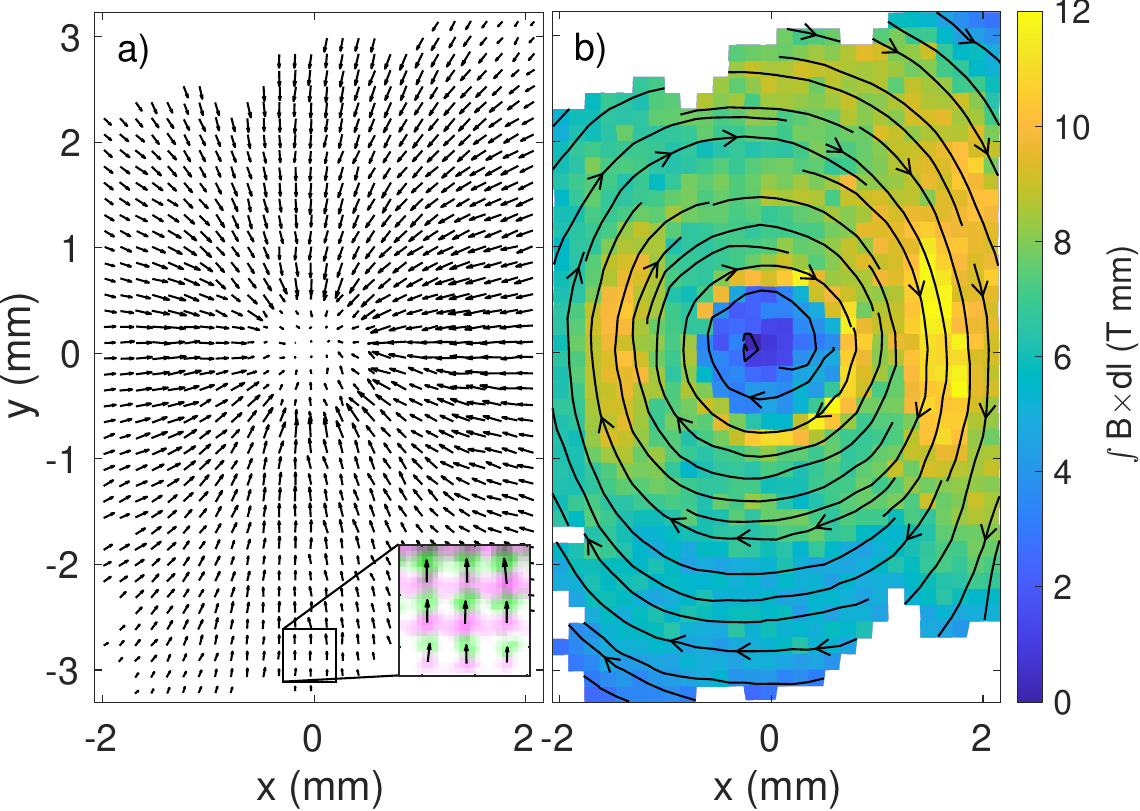}
    \caption{\textbf{Deflection map and B field.} a) Deflection from x-ray to 14.7 MeV proton beamlet position in the full mesh configuration. Inset: Overlaid IP and CR-39 images showing beamlet deflection from x-rays (pink) to protons (green) outside of the nominal plasma bubble. b) Path-integrated magnetic field with field streamlines.}
    \label{fig:DeflectionMap_Bth}
\end{figure}

We achieve high fidelity magnetic inversions of Biermann-battery magnetic fields with anomalously large magnetic extent and flux. Figure \ref{fig:DeflectionMap_Bth}a shows the 14.7 MeV proton deflection map for the full-mesh configuration. The map has been rotated 18 degrees from Figures \ref{fig:Combined_IP_CR-39}a,d so that the rows in the x-ray mesh are horizontal. The deflection is predominantly inward corresponding to a clockwise toroidal magnetic field. Notably, deflection is observed out to the boundaries of the detector, which might go unnoticed without direct comparison to the x-ray beamlets (see e.g. inset in Figure \ref{fig:DeflectionMap_Bth}a). The relative position of the teeth compared to the edge of the targets in Figure \ref{fig:Combined_IP_CR-39} is another clear and qualitative indication that there is significant proton deflection all the way to the edge of the field of view, implying fields at large radius. Proton deflection at the edge of the target is dominated by magnetic fields and the electric contribution is shown to be non-significant. A detailed analysis using 3 and 14.7 MeV protons to distinguish between E and B fields is presented in Section \ref{sec:EvB} below.

Figure \ref{fig:DeflectionMap_Bth}b shows the path-integrated magnetic field where two rings of increased field strength are visible. The first ring is peaked at $r=0.7$ mm and produces the Biermann ``ring'' feature in the proton image of Figure \ref{fig:Combined_IP_CR-39}e. The second ring is visible at $r=1.7$ mm and corresponds to the location where the field vanishes in several prior experiments and the edge of the expanding plasma bubble in their supporting MHD simulations \cite{li2007observation,petrasso2009lorentz,gao2015precision,campbell2022measuring}. The measured field profile qualitatively matches prior work inside the bubble in terms of maximum path-integrated field strength and where that maximum occurs. In contrast to prior work, the measured fields extend beyond the bubble boundary and out to the edge of the mesh, which has a maximum radius of 3.5 mm.

\begin{figure}
    \includegraphics[width=\linewidth]{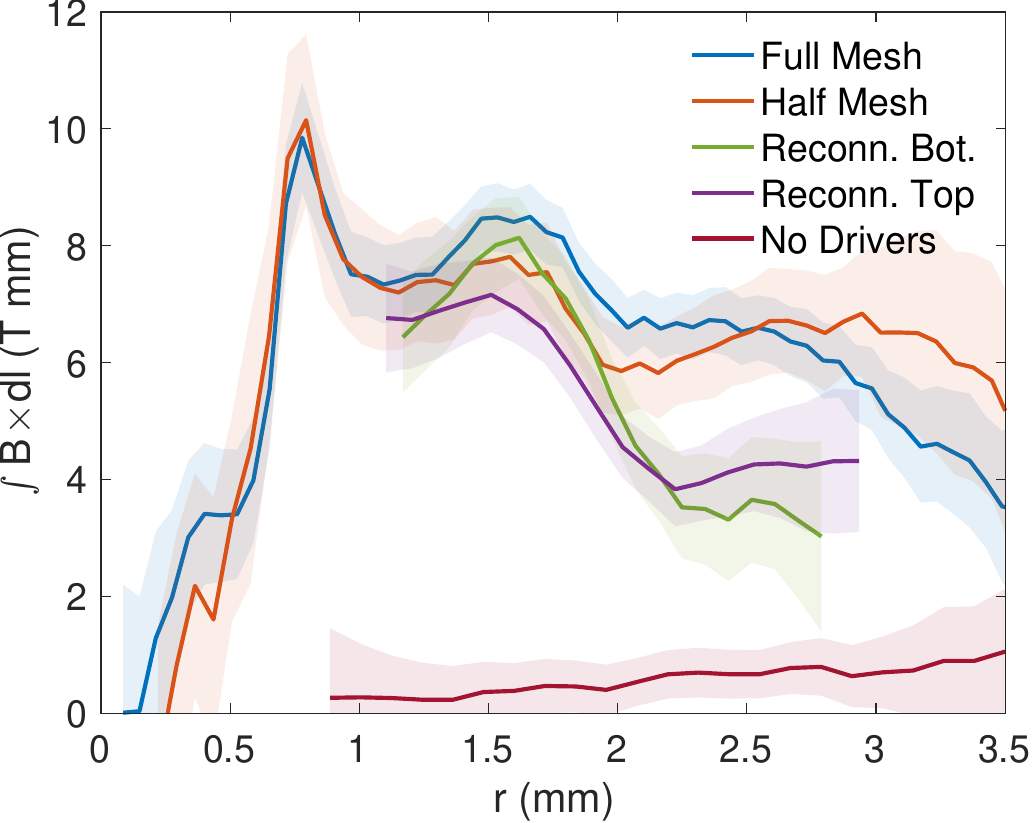}
    \caption{\textbf{Magnetic field profile} 
    Radial profile of the path-integrated toroidal magnetic field after averaging azimuthally. The shaded region represents the uncertainty in field strength in each radial bin, accounting for uncertainties in image alignment, beamlet selection, and electric field contribution. In the reconnection geometry, the bottom and top regions are treated independently. 
    Non-paraxial corrections to the path-integrated field strength are included as discussed in Appendix A.}
    \label{fig:Bth_combined}
\end{figure}

Magnetic fields from the other experimental configurations show good agreement with the full mesh case and also suggest the presence of far-reaching magnetic fields that extend past the plasma bubble. Figure \ref{fig:Bth_combined} shows the radial profile of the toroidal magnetic field after averaging azimuthally. The path-integrated field consistently peaks around the edge of the bubble and extends past it to over 3.5 mm in radius. 

The lower bound on the magnetic flux is calculated as $\Phi_B=\int dz\,dr\,B_\theta=$ 21 T$\,$mm$^2$ for full and half mesh configurations (we emphasize this is a lower bound because we do not observe the field go to zero within the field of view). These values are over three times higher than similar experiments that did not use x-ray fiducials \cite{campbell2022measuring,li2007observation,gao2015precision,petrasso2009lorentz}. Non-paraxial geometrical corrections are included in these calculations as discussed in Appendix A. Table \ref{tab:CompTable} shows the core experimental parameters and results of prior experiments in comparison to this work. Under a similar experimental regime as prior experiments, this work measures strikingly different magnetic flux and field extent.
\begin{table*}
    \centering
    \begin{tabular}{c|ccc|cccc}
        Source 
        & \multicolumn{1}{p{2cm}}{\centering Laser Energy\\ (kJ)}
        & \multicolumn{1}{p{3cm}}{\centering Peak Intensity\\$(10^{14}$ W$/$cm$^{2}$ )}
        & \multicolumn{1}{p{2cm}}{\centering Probe Time \\(ns)}
        & \multicolumn{1}{|p{2.5cm}}{\centering $(\int B\times dl)_{max}$ \\ (T$\,$mm)}
        & \multicolumn{1}{p{1.75cm}}{\centering $r(B_{max})$ \\ (mm)}
        & \multicolumn{1}{p{2cm}}{\centering $\Phi_B$ \\ (T$\,$mm$^2$)}
        & \multicolumn{1}{p{1.75cm}}{\centering $r_{outer}$ \\ (mm)} \\
        \hline
        
        \cite{li2007observation}& 0.5 & 1 & 1.5 & 11 & --- & --- & 1\\
        \cite{petrasso2009lorentz}& 0.5 & 0.8 & 1.36 & 10 & 0.75 & 7 & 1.4\\
        \cite{gao2015precision}& 2 & 4 & 1.2 & --- & --- & --- & 1.6\\
        \cite{campbell2022measuring}& 4 & 4.4 & 1.2 & 9 & 0.9 & 4.5 & 1.5\\
        \hline
        This work& 1 & 3 & 1.4 & 10 & 0.7 & 21 & $>$3.5\\
    \end{tabular}
    \caption{\textbf{Comparison to prior work.} Note: $r_{outer}$ is defined as the radius where the path-integrated magnetic field falls to 10\% of its maximum value.}
    \label{tab:CompTable}
\end{table*}

The control shot without drive beams still has the signature of a weak toroidal field after averaging azimuthally, albeit much less than the other field measurements. However, an analysis using 3 MeV protons reveals that the control shot has the \textit{opposite} magnetic polarity as the plasma shots with strength 2 T$\,$mm which is at the edge of the measurement uncertainty. Instead, the inward deflection is due to a radial electric field of strength 100 to 200 MV$\,$m$^{-1}\,$mm. The counter-clockwise magnetic field orientation is consistent with an electron current from the backlighter to the target. Additional analyses of the E fields in other configurations demonstrates that E fields are subdominant and have inconsistent orientation, as shown in the next section.

\section{E vs. B Field Analysis} \label{sec:EvB}

\begin{figure} [b]
    \includegraphics[width=0.9\linewidth]{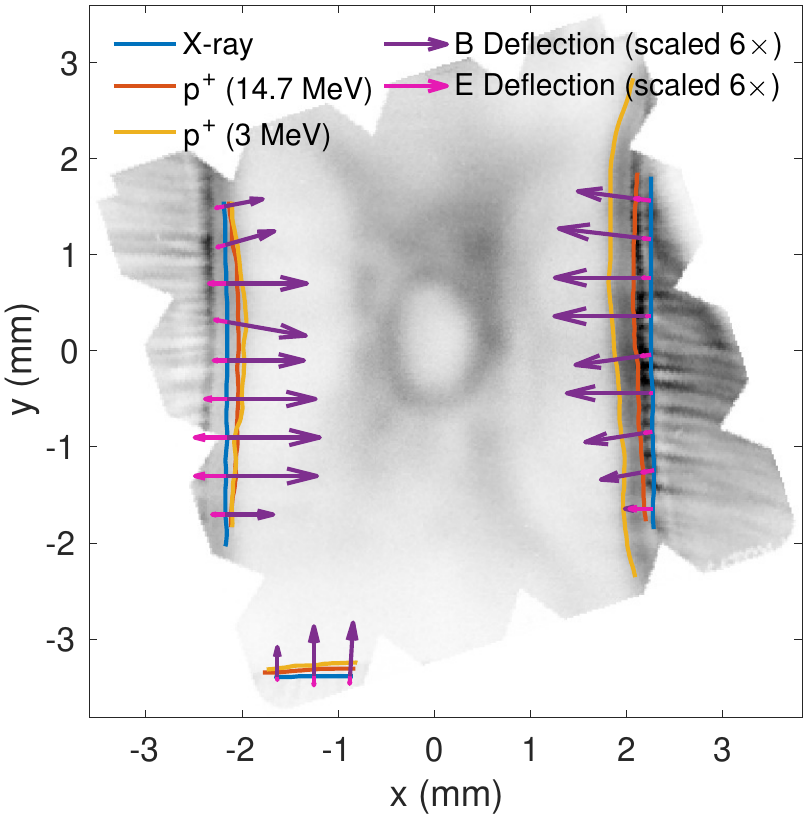}
    \caption{\textbf{3 MeV deflection analysis.} 3 MeV proton flux for the full mesh configuration with overlaid edges of the target for x-rays (blue), 14.7 MeV protons (red), and 3 MeV protons (yellow). The contribution of the 14.7 MeV proton deflection from magnetic (purple) and electric (pink) fields is shown with vector magnitude scaled up by 6$\times$ for visibility.}
    \label{fig:3MeVDeflection}
\end{figure}

3 MeV protons were also produced by the fusion backlighter and simultaneously measured on CR-39. Comparing deflections between 14.7 and 3 MeV protons discriminates between electric and magnetic fields, since the deflection from magnetic fields scales with proton energy as $d_B\sim K^{-1/2}$ and from electric fields as $d_E\sim K^{-1}$. The total deflection of both proton populations can be written as a combination of electric and magnetic field contributions as
\begin{equation}
    d_{14.7}=d_E+d_B,\ d_3=Rd_E+\sqrt{R}d_B
\end{equation}
where $d_{14.7}$ and $d_3$ are the deflection of 14.7 and 3 MeV protons, respectively, and $R=14.7/3$ is the kinetic energy ratio of proton populations. The system is solved for $d_E$ and $d_B$ and the path-integrated field strengths are extracted. In this analysis, we assume the 200 ps difference in time of flight between the two proton energies is negligible when compared to the field evolution time-scale of $\sim$1 ns.

Figure \ref{fig:3MeVDeflection} shows the 3 MeV proton flux in the full mesh configuration with overlaid target edges for the x-rays, 3 MeV protons, and 14.7 MeV protons. Sharp features like the 3 MeV beamlets were blurred out due to proton scattering, but larger features like the edges and corners of the target are still visible. 
Deflections of the target edge are only sensitive in the direction perpendicular to the edge. Thus, a deflection analysis only extracts the perpendicular component of the electric field and the parallel component of the magnetic field. 
These components should approximate the full field at the equator and poles of the interaction where the perpendicular direction coincides with the radial direction. Magnetic fields are confirmed as the dominant source of beamlet deflection at the edge of the target. A path-integrated electric field at the threshold of the measurement uncertainty (100 MV$\,$m$^{-1}\,$mm) is detected on both edges which is small relative to the observed deflection. This electric field is consistently oriented to the left (see Fig. \ref{fig:3MeVDeflection}) and could be generated from the target and mesh charging from high-energy backlighter products. Analyses of the other mesh configurations also support magnetic fields extending out to the edge of the target.
\section{2D Gorgon Simulations}

Experimental results were compared to 2D, cylindrically symmetric simulations using the extended MHD code Gorgon \cite{walsh2017self,walsh2020extended,chittenden2004x,ciardi2007evolution} and found to differ significantly. Simulations include Biermann-battery field generation, Nernst advection \cite{nishiguchi1985nernst}, Righi-Leduc heat flux, and radiation transport. Improved magnetic transport coefficients were used \cite{sadler2021symmetric,davies2021transport}, which have been shown to be important in the low magnetization regime relevant to this work \cite{walsh2021updated}. In addition, several models of Biermann-battery suppression are incorporated to account for nonlocal field generation in the hot, rarefied corona where kinetic effects become important \cite{davies2023nonlocal,sherlock2020suppression}. 

\begin{figure} [b]
   \includegraphics[width=\linewidth]{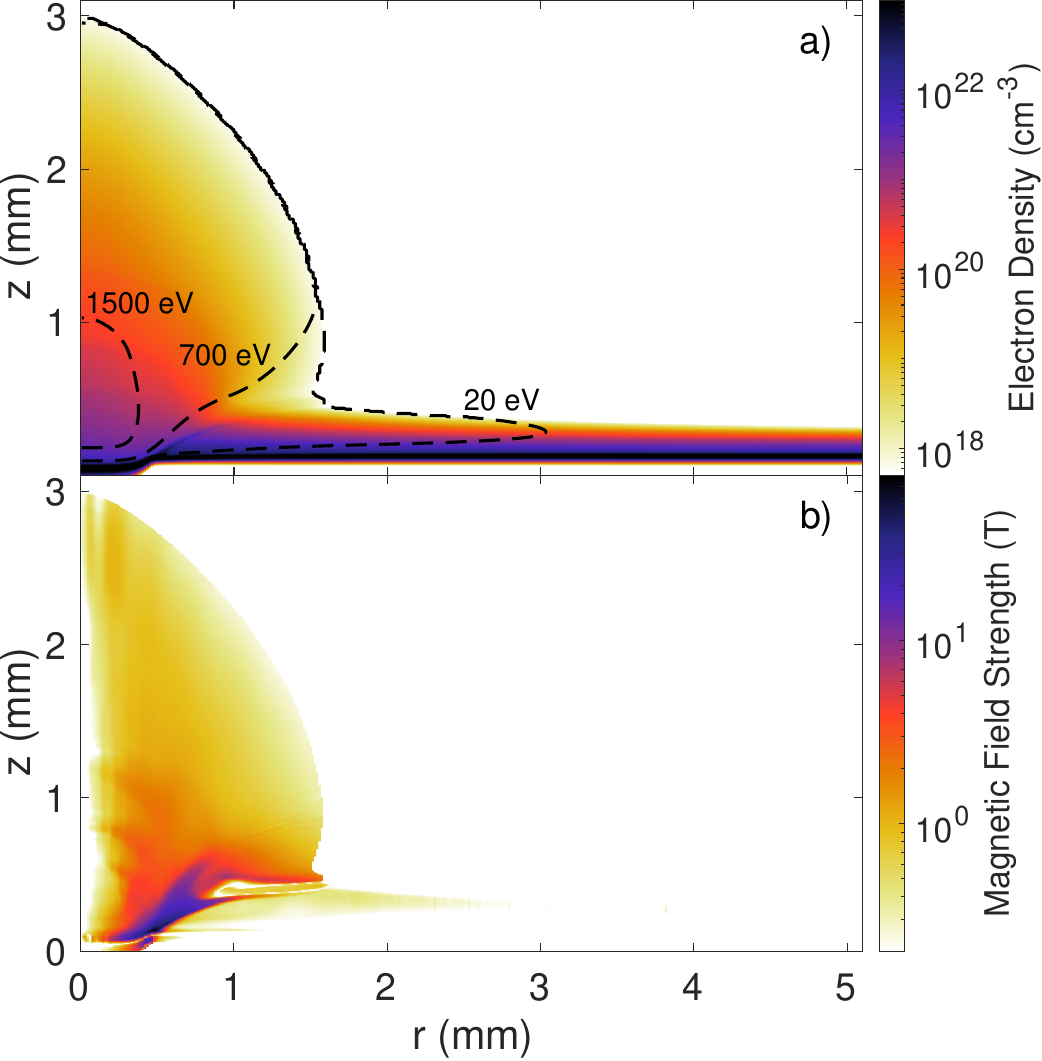}
    \caption{\textbf{Gorgon simulation results}. 
    a) Electron density with overlaid lines of constant electron temperature at $t=1.4$ ns. b) Azimuthal magnetic field strength. The field sign is predominantly negative (out of the page), although there are small patches of positive field (e.g. below the target surface). Simulations use the Sherlock Biermann-battery suppression model \cite{sherlock2020suppression}.}
    \label{fig:nBT_Gorgon}
\end{figure}

Figure \ref{fig:nBT_Gorgon} shows the electron density and azimuthal magnetic field strength from Gorgon simulations using the Sherlock \cite{sherlock2020suppression} suppression model at $t=1.4$ ns. The overlaid dashed lines in Figure \ref{fig:nBT_Gorgon}a are contours of constant electron temperature. An expanding bubble of plasma is visible and demarcated by the sharp decline in electron temperature from $700$ to $20$ eV. The magnetic field generation is primarily within this bubble, with the greatest field strength occurring close to the surface of the foil near the edge of the laser spot. Some weak fields are generated along the surface of the target as seen in Gao \emph{et al.} \cite{gao2015precision}, however these are at the $<1$ T level and are negligible when path-integrated. The other Biermann-suppression schemes yield similar field structure and strengths. 

\begin{figure} [t]
    \includegraphics[width=\linewidth]{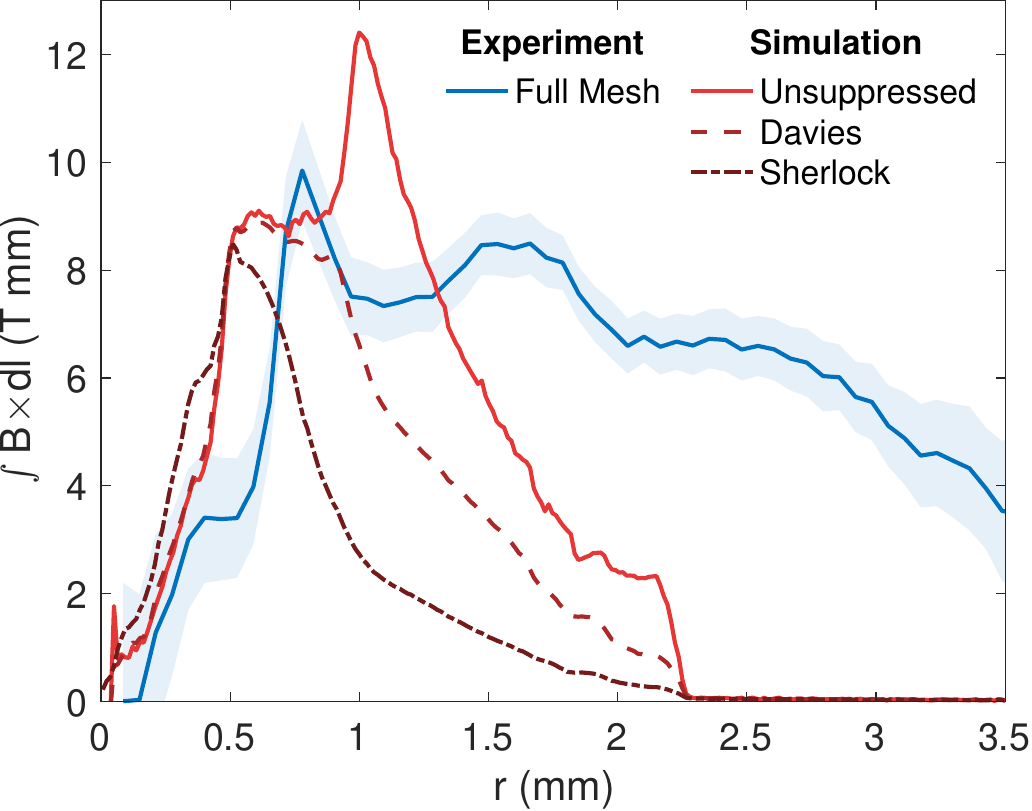}
    \caption{\textbf{Magnetic field comparison to simulations}. 
    Radial profiles of the path-integrated toroidal magnetic field, accounting for non-paraxial path integration. Experimental results in the full mesh configuration are shown in blue as repeated from  Figure \ref{fig:Bth_combined}. Profiles from Gorgon simulations are shown in shades of red. Different Biermann-battery suppression models are used including no suppression (solid line), Davies model (dashed line) \cite{davies2023nonlocal}, and Sherlock model (dot-dashed line) \cite{sherlock2020suppression}. Field integration accounts for non-paraxial geometry.}
    \label{fig:Bth_combined_walsh}
\end{figure}

Figure \ref{fig:Bth_combined_walsh} shows the path-integrated magnetic field profiles in experiment and simulations. The profile from the full mesh experimental configuration (solid blue line) is repeated from Figure \ref{fig:Bth_combined}. Gorgon simulations with different Biermann-battery suppression models are plotted in comparison (red lines); the solid line had no suppression and the dashed and dot-dashed lines used suppression models from Davies \emph{et al.} \cite{davies2023nonlocal} and Sherlock \emph{et al.} \cite{sherlock2020suppression}, respectively. Note that the simulation results using Davies' suppression model do not include the suggested corrections by magnetization; when these were included, the unsuppressed case and the Davies case were indistinguishable. The different suppression models yield different path-integrated magnetic field profiles, however the field extent is consistently limited to the bubble radius of $\sim2$ mm. Prior experimental and simulation field profiles \cite{campbell2022measuring,gao2015precision,petrasso2009lorentz,li2007observation} are consistent with the Davies or Sherlock simulations. In prior work, the fields rise to a single peak of order $\sim$10 T$\, $mm before vanishing at the bubble edge. 

Although good agreement is found between our experiment results and simulations close to the laser axis ($r<0.7$ mm), the experimental field extends beyond simulations by a factor of $\gtrsim2\times$.
The disagreement between cutting-edge simulations and experiment suggests that the Biermann-battery effect in the extended MHD framework cannot fully explain the results presented here. In the future, particle-in-cell simulations will be performed to explore non-MHD effects.

\section{Discussion}
We may also estimate the energy contained in the magnetic fields to provide a further point of comparison to prior experimental results. As a simplifying assumption, the field is taken to be localized to a slab of thickness $L_B$ in the out-of-plane direction. The magnetic energy can then be estimated as $U_B=1/(2 \mu_0)\int dV B^2 \sim \pi / (\mu_0 L_B) \int dr\, r\, (B\,dl)^2$ assuming azimuthal symmetry and field strength purely a function of radius $B(r)$. 
In order to infer the field energy, we estimate the field thickness as the temperature gradient length-scale as presented in prior work: $L_B=T_e/\nabla T_e \sim$ 100 $\mu$m \cite{li2006measuring,li2007observation}. This implies a maximum field strength of $B_{max} \sim 100$ T and lower bound of the magnetic energy $\sim$ 7.5 J. In contrast, the magnetic energies of both \cite{petrasso2009lorentz} and \cite{campbell2022measuring} are estimated to be $\sim1.5$ J using the published magnetic field profile at the same thickness. Within this simplified geometry, the inferred magnetic energy presented here is over 5$\times$ larger than prior works. (Note that the ratio of magnetic energy in this work compared to that inferred from \cite{petrasso2009lorentz,campbell2022measuring} is independent of $L_B$ since the same thickness is assumed for both calculations).

We do not yet have a complete model for the large field extent and magnetic flux, but the measurement presented here is novel and important. One likely component in the narrative is that there is plasma outside of the nominal plasma bubble. In order to support toroidal fields at large radius, plasma is needed to close the poloidal current loop. In addition, the difference in magnetic flux measurement may suggest plasma at large $z$. As shown in Walsh \emph{et al.} \cite{walsh2021biermann} the generation of magnetic flux due to the Biermann-battery effect can be rewritten as an integral solely along the laser axis 
\begin{equation}
    \left[\frac{\partial \Phi_B}{\partial t}\right]_{Biermann} = \int dz\, \frac{\nabla P_e}{en_e}.
    \label{eqn:Flux_axis}
\end{equation}
Therefore, the elevated magnetic flux observed here should be derivable from the on-axis plasma parameters. These two observations are likely coupled and may suggest a low-density plasma 'pre-blowoff' from the foil, both at large $r$ and at large $z$. 

The difference in magnetic flux  suggests that there is not only fast advection of fields to large radius, but also more field generation in general. For example, the results here cannot be explained simply by Nernst advection \cite{nishiguchi1985nernst}. Although the direction of Nernst advection is radially outwards (down temperature gradients away from the laser spot) which is consistent with the extended field measurement, the Nernst effect is a conservative term in the induction equation and cannot generate magnetic flux or explain the elevated flux in this work. Still, it is expected to be important for field advection outside of the focal region at early times \cite{willingale2010fast}. There may also be non-MHD effects like fast electrons as seen in simulations in \cite{willingale2010fast} and separation of ion species that could lead to additional magnetic field generation mechanisms. For example, Carbon and Hydrogen ions have different sound speeds and may naturally separate in their expansion and contribute to field generation at their interface.

Magnetic fields can significantly alter the physics inside of an ICF implosion. Most notably, electron heat conduction is suppressed in directions perpendicular to the field \cite{froula2007quenching}. Recent experiments at the NIF have demonstrated enhancements of fusion yield and ion temperature when an axial magnetic field is applied to the hohlraum \cite{moody2022increased}. Extended MHD simulations \cite{farmer2017simulation,walsh2017self} also show the presence of self-generated magnetic fields in hohlraums. These too can magnetize particles and modify heat transport. The results presented here show extended magnetic field generation which is underestimated by simulations. This suggests simulations may also underestimate the extent of magnetic field generation in hohlraums, especially in the low density, coronal plasma where our experimental results diverged from simulations.



\section{Conclusion}
We report the first measurement of Biermann-battery magnetic fields using a proton radiography scheme with \textit{in situ} x-ray fiducials. 
The magnetic field measured here extends past the expanding plasma bubble to $r>3.5$ mm by $t=+1.4$ ns. Under a similar experimental regime as prior work, we report significantly further field extent and greater magnetic flux owing to the use of x-ray fiducials.
This radiography scheme is demonstrated as a valuable technique for high fidelity mesh-based field reconstructions. The reference pattern of the mesh imprinted on the x-ray image enables measurements of electromagnetic fields in systems where the fields extend beyond the radiograph field of view and can be used to both aid and benchmark traditional inversion techniques \cite{bott2017proton}. The deviation from prior experimental and simulation results suggests a new physics mechanism for the magnetic field generation and transport in laser-solid interactions. A larger spatial extent of magnetic fields and larger energy stored in magnetic fields would significantly impact hohlraum plasma physics by magnetizing particles and modifying plasma heat transport. Follow-on experiments have been conducted that investigate the 3D structure of the magnetic field (i.e. shape and height off of the target surface) through tomographic proton radiography that will further inform the relevant physics and help to constrain additional simulations.

\section*{Acknowledgements}
The authors thank the OMEGA staff for their help in conducting these experiments and General Atomics for target fabrication. This work was supported by the Department of Energy under grant No. DE-NA0004034.

\section*{Appendix A: Geometric Effects}
\renewcommand{\thefigure}{A\arabic{figure}} 
\renewcommand{\theequation}{A\arabic{equation}} 

\setcounter{figure}{0}
\setcounter{equation}{0}

\begin{figure} [t]
    \includegraphics[width=\linewidth]{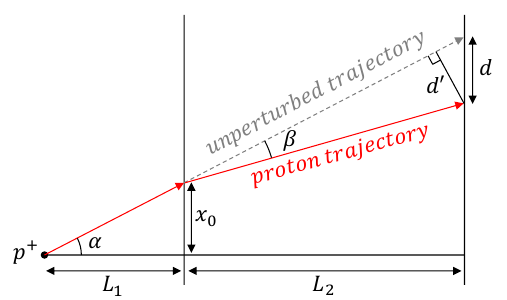}
    \caption{\textbf{Non-paraxial geometry.} Illustration of the geometry for non-paraxial corrections.}
    \label{fig:paraxial}
\end{figure}

Calculating the path-integrated magnetic field from beamlet deflections in Eq. (\ref{eqn:beamlet_def}) relies on the smallness of the paraxial parameter, here defined to be the ratio between the transverse magnetic field extent and the source-to-plasma length. However, this approximation begins to break down at the edges of the image due to the large extent of magnetic fields in this work. The path-integrated field strength is correspondingly reduced by the geometrical factor $\cos^2{\alpha}$ where $\alpha$ is the proton emission angle from target normal (see Figure \ref{fig:paraxial}). After accounting for non-paraxial corrections, the path-integrated magnetic field is calculated as

\begin{equation}
    \label{eqn:beamlet_def_parax}
    \int d\mathbf{l} \times \mathbf{B} = \frac{m_p v_z \mathbf{d}}{e L_2} \cos^2\alpha.
\end{equation}

An additional factor of $\cos{\alpha}$ is also included for magnetic flux calculations to account for geometrical effects in the area integral: $\int dr\,dz \approx \int dr\,dl\, \cos \alpha$. For the furthest beamlets at $\alpha=20$\degree, the path-integrated fields are reduced by 12\% and the local flux is reduced by 17\%. This correction is reflected in the magnetic field profile (Fig. \ref{fig:Bth_combined}) and in magnetic flux and energy calculations. Note that Eq. (\ref{eqn:beamlet_def_parax}) still assumes small deflection angle, $\beta$.

The cone geometry of the proton source may contribute slightly to the large apparent extent of the fields. If the magnetic fields lie above the surface of the foil, as they do in MHD simulations \cite{campbell2022measuring,li2006measuring}, then the fields are projected onto larger radii in the plane of the foil. For a magnetic field with mean height above the surface $h$, the field extent is scaled by $h/(L_1-h)$. Although this projection may overestimate the field extent at certain locations (e.g. 25\% further for a field height of 2 mm), simulations suggest the far reaching fields are close to the surface of the target \cite{campbell2022measuring,li2006measuring,gao2015precision} which would be unaffected by the projection geometry. Furthermore, if we attempt to attribute the extended field entirely to the projection geometry while staying within the nominal bubble radius of $r=1.7$ mm, we find that the field must be at an ``average height" of 5 mm. This corresponds to the physically unreasonable assertion that the fields totally fill the volume between the foil and the capsule (10 mm away). Finally, the effect of the projection geometry was not observed in \cite{petrasso2009lorentz}, which would have manifested as a large asymmetry between the two bubbles. As such, this correction does not change the main result that the fields extend beyond the plasma bubble and much further than previously measured.

\bibliography{BiermannBatteryPRadBib}

\end{document}